\documentclass[useAMS,usenatbib]{mn2e}

\usepackage{times}
\usepackage{multirow}

\usepackage{psfrag}
\usepackage{pspicture}
\usepackage{graphicx}
\usepackage{amsmath}
\usepackage{color}

\title[Spectral age modelling of the `Sausage' cluster radio relic]{Spectral age modelling of the `Sausage' cluster radio relic}
\author[A. Stroe et al.]{Andra Stroe$^{1}$\thanks{E-mail: astroe@strw.leidenuniv.nl}, Jeremy J. Harwood$^{2}$, Martin J. Hardcastle$^{2}$, Huub J. A. R\"ottgering$^{1}$\\
$^{1}$Leiden Observatory, Leiden University, P.O.\ Box 9513, NL-2300 RA Leiden, The Netherlands\\
$^{2}$School of Physics, Astronomy and Mathematics, University of Hertfordshire, College Lane, Hatfield, Hertfordshire AL10 9AB, UK\\}

\begin{document}
\maketitle

\begin{abstract}
CIZA J2242.8+5301 is a post-core passage, binary merging cluster that hosts a large, thin, arc-like radio relic, nicknamed the `Sausage', tracing a relatively strong shock front. We perform spatially-resolved spectral fitting to the available radio data for this radio relic, using a variety of spectral ageing models, with the aim of finding a consistent set of parameters for the shock and radio plasma. We determine an injection index of $0.77^{+0.03}_{-0.02}$ for the relic plasma, significantly steeper than was found before. Standard particle acceleration at the shock front implies a Mach number $M=2.90^{+0.10}_{-0.13}$, which now matches X-ray measurements. The shock advance speed is $v_\mathrm{shock}\approx2500$ km s$^{-1}$, which places the core passage of the two subclusters $0.6-0.8$ Gyr ago. We find a systematic spectral age increase from $0$ at the northern side of the relic up to $\sim60$ Myr at $\sim145$ kpc into the downstream area, assuming a $0.6$ nT magnetic field. Under the assumption of freely-ageing electrons after acceleration by the `Sausage' shock, the spectral ages are hard to reconcile with the shock speed derived from X-ray and radio observations. Re-acceleration or unusually efficient transport of particle in the downstream area and line-of-sight mixing could help explain the systematically low spectral ages.
\end{abstract}

\begin{keywords}
acceleration of particles -- shock waves -- radiation mechanisms: non-thermal -- radio continuum: individual: CIZA J2242.8+5301 -- galaxies: clusters: intracluster medium
\end{keywords}
     
\section{Introduction}
\label{sec:intro}
\subsection{Cluster radio relics}
Acceleration of particles at shock fronts is an ubiquitous phenomenon in astrophysical contexts \citep[e.g.][]{2012SSRv..166...71B}. In the presence of magnetic fields, particles accelerated to relativistic speeds generate synchrotron radiation. Most of the observed radio emission in objects ranging from supernova remnants (SNR) to active galaxies and clusters of galaxies is due to synchrotron processes \citep{1979rpa..book.....R}. Magnetic field strengths, shock strengths and spatial scales vary enormously between these different object classes. Shock fronts in SNR have Mach numbers of the order of $1000$, but are limited in size to a couple of parsecs \citep[e.g.;][]{2008ARA&A..46...89R}. In radio galaxies and quasars, particle transport occurs via a jet terminating in a $M\sim10-100$, kpc-long shock \citep{1984ARA&A..22..319B}. Some of the weakest (Mach number of a few), but most extended (Mpc-wide) observable astrophysical shocks are found in the intracluster medium (ICM) of clusters \citep[see reviews by][]{2012SSRv..166..187B, 2012A&ARv..20...54F, 2014IJMPD..2330007B}. Galaxy clusters thus offer us the possibility to study the physics of weak shocks. 

Since most of their observable mass is in the form of diffuse gas located in between the galaxies, clusters possess a huge reservoir of particles that could radiate synchrotron emission in the presence of magnetic fields and an accelerating shock. Shocks are expected given the violent merger history undergone by massive clusters within hierarchical structure formation \citep{2002ASSL..272....1S}. Relatively strong (a few $10^{-10}$T) magnetic fields can be found out to the outskirts of clusters which, in the presence of shocks, enable the production of synchrotron emission \citep[e.g.][]{2009A&A...503..707B, 2010A&A...513A..30B}. 

Direct evidence for shock-accelerated particles has been found at the periphery of approximately $40$ merging clusters in the form of diffuse, extended, polarized, synchrotron emission \citep{2012A&ARv..20...54F}. These objects are known as giant radio relics. Sensitive observations show that, in the case of well behaved, non-disrupted, arc-like shocks, the spectral index of the radio emission becomes steeper from the shock into the downstream area. Examples include Abell 2744 \citep{2007A&A...467..943O}, Abell 521 \citep{2008A&A...486..347G}, CIZA J2242.8+5301 \citep{2010Sci...330..347V, 2013A&A...555A.110S} and 1RXS J0603.3+4214 \citep{2011MNRAS.418..230V}. The particles are thought to be accelerated via diffusive shock acceleration \citep[DSA;][]{1983RPPh...46..973D} to a power law energy distribution at the shock front, which results in a power-law radio spectrum (with the slope called the `injection index'). With time, under the effect of radiative losses which are proportional to the energy of the particle squared, the higher energy electrons preferentially cool, leading to a steepening of the spectrum at the higher frequencies \citep{1979rpa..book.....R}. 

\subsection{The `Sausage' cluster}
CIZA J2242.8+5301 is a cluster that hosts a varied collection of bright, diffuse sources, including twin, double radio relics and smaller, patchy extended sources \citep[for source labelling see][] {2013A&A...555A.110S}. The peculiar morphology of the northern relic garnered the nickname the `Sausage': the source has an arc-like shape, thought to trace a spherical shock, seen in projection, moving towards the north. \citet{2010Sci...330..347V} performed a detailed study of the `Sausage' relic and discovered steepening of the spectral index from the front towards the back of the shock (north to south). They concluded that the source, at least in proximity to the shock front, is permeated by a coherent, aligned, strong magnetic field of $0.5-0.7$ nT. On the basis of spectral index and curvature maps and colour-colour plots, \citet{2013A&A...555A.110S} concluded that the observed emission at each location within the relic is well described by a single injection spectrum with subsequent spectral ageing. The spectral analysis suggested that the outer edge of the northern relic is traced by a shock of Mach number $4.58\pm1.09$. The plasma trailing behind the shock shows systematic steepening of the spectrum further into the downstream area, indicative of spectral ageing. X-ray observations from the \textit{XMM-Newton}, \textit{Suzaku} and \textit{Chandra} telescopes indicate the presence of weak discontinuities coinciding with the location of the radio shocks \citep{2013PASJ...65...16A, 2013MNRAS.429.2617O, 2013MNRAS.433..812O, 2014MNRAS.440.3416O}. These discontinuities indicate a much weaker shock strength than the radio: $M\sim2.5$. The shocks were also found to possibly interact with the cluster galaxies, boosting the H$\alpha$ luminosity function of galaxies near the radio relics \citep{2014MNRAS.438.1377S}, leading to enhanced numbers of bright star-forming galaxies when compared to blank fields.

\subsection{Main questions and aim of this paper}
The general properties of relics have been broadly measured, but there are many open questions that have to be addressed \citep[see review by][]{2014IJMPD..2330007B}. Does the relationship often expected between the injection $\alpha_\mathrm{inj}$ and integrated index at high radio frequencies hold $\alpha_\mathrm{int}$ ($\alpha_\mathrm{int}-\alpha_\mathrm{inj}=0.5$, where $F_{\nu}\propto\nu^{-\alpha}$) hold \citep[e.g.][]{pacholczyk, 2002MNRAS.337..199M}, given that the high-frequency integrated index of relics varies depending on the frequency window \citep{AMI}? Is DSA efficient enough at low Mach numbers to inject/accelerate enough electrons to explain the total radio emission \citep[e.g.][]{2007ApJ...669..729K,2012ApJ...756...97K}? How can we explain the discrepancy between the radio and X-ray detected shocks, where the X-ray Mach number is systematically lower than the radio one \citep{2013PASJ...65...16A, 2013MNRAS.429.2617O, 2013MNRAS.433..812O, 2014MNRAS.440.3416O}? How are strong magnetic field generated within radio relics? Why is the surface brightness so uniform along some radio relics?

To address some of these questions we need to measure physical parameters such as the injection index, Mach number and spectral age precisely. The `Sausage' relic is a bright, well resolved cluster relic, with excellent radio data available, which enables us to test spectral ageing models. In this paper we derive a consistent set of well understood physical parameters using some of the highest quality radio data available for any radio relic. 

We present a spectral ageing analysis of the `Sausage' relic using six radio frequency maps from the Giant Metrewave Radio Telescope (GMRT) and the Westerbork Synthesis Radio Telescope (GMRT) presented by \citet{2013A&A...555A.110S}, and references therein. Using the Broadband Radio Astronomy ToolS \citep[{\sc brats}\footnote{http://www.askanastronomer.co.uk/brats/};][]{2013MNRAS.435.3353H}, we fit spectral ageing models at the pixel-by-pixel level, with the aim of determining the physical model that best describes the spectral properties and morphology of the source. In order to do so, we test the Jaffe-Perola \citep[JP;][]{1973A&A....26..423J} and the Kardashev-Pacholczyk \citep[KP;][]{1962SvA.....6..317K, pacholczyk} models of spectral ageing, which assume a single burst of electron injection, followed by radiative losses in a constant magnetic field. While the KP model assumes that the pitch angle between the electrons and the magnetic field stays constant over time, the JP model is more realistic in that it allows for the isotropisation of the pitch angle. We also test the more complex Tribble model, which accounts for emission arising from the more physically plausible case of an electron population subject to a spatially inhomogeneous magnetic field \citep{1993MNRAS.261...57T, hardcastle2013, 2013MNRAS.435.3353H}. For the first time we derive precise injection indices ($\alpha_\mathrm{inj}$), the Mach numbers ($M$) and the ages of the electrons ($t_\mathrm{age}$, time since last acceleration) in a self-consistent way using model fits to many different regions of the relic. 

The structure of this paper is the following: in Section~\ref{sec:obs-reduction} we give an overview of the observations, in Section~\ref{sec:spec} we present the parameters and models tested with {\sc brats} and in Section~\ref{simulations} we discuss the effect of finite resolution on our results. In Section~\ref{sec:discussion} we discuss the results of the model fitting and the derived physical properties and  the implications for cluster relics and ICM properties. The main results are summarised in Section~\ref{sec:conclusion}. We assume a flat, $\Lambda$CDM cosmology with $H_{0}=70.5$~km~s$^{-1}$~Mpc$^{-1}$, matter density $\Omega_M=0.27$ and dark energy density $\Omega_{\Lambda}=0.73$ \citep{2009ApJS..180..306D}. At the `Sausage' redshift of $0.19$, $1$~arcmin on the sky measures $0.191$~Mpc. All images are in the J2000 coordinate system.

\begin{table}
\begin{center}
\caption{Sensitivity of the images used for the analysis of the `Sausage' relic.}
\begin{tabular}{c c c}
\hline
\hline
Frequency & RMS  & Telescope \\
   (MHz)  & $\mu$Jy beam$^{-1}$ &  \\ \hline
153 & 1800 & GMRT \\
323 &  240 & GMRT \\
608 &   48 & GMRT \\
1221 &  58 & WSRT \\
1382 &  20 & WSRT \\
1714 &  33 & WSRT \\
\hline
\end{tabular}
\label{tab:images}
\end{center}
\end{table}

\section{Observations and imaging}\label{sec:obs-reduction}
The `Sausage' cluster was observed at eight radio frequencies spanning between $153$~MHz and $16$~GHz with the GMRT, WSRT and the Arcminute Microkelvin Imager telescopes between 2009 and 2013. A full description of the observations and data reduction is given by \citet{2013A&A...555A.110S} and \citet{AMI}. The flux calibration error on these maps was taken to be $10$ per cent.

The $16$ GHz AMI data of the relic published by \citet{AMI} is not suitable for fitting of spectral ageing models on a pixel-by-pixel basis, as the relic downstream area is not resolved. Hence the $16$ GHz is not used the following analysis.

The WSRT measurements were taken in a single configuration. Due to the scaling of the angular sizes sampled by the same telescope configuration at different frequencies, the $2272$~MHz data has the sparsest inner uv-coverage.This effectively means that the interferometer might be resolving out some parts of the largest scale diffuse emission. The brightness distribution of the $2272$~MHz map is dominated by small-scale features ($\sim30$~kpc), an effect which is not seen in the lower frequency maps. We attribute this to the poor uv-coverage. In addition, because of the steep spectrum of the relic and the enhanced noise levels, the S/N in the $2272$~MHz map is poor. Therefore, we prefer to map as much of the diffuse emission as possible using six radio frequencies and remove the $2272$ MHz map from further analysis. 

To ensure recovery of flux on the same spatial scales for the GMRT ($153$, $323$, $608$ MHz) and WSRT ($1221$, $1382$, $1714$ MHz) observations, we kept data on common uv-distances for all the remaining six frequencies. Data taken below $0.2$ k$\lambda$ was discarded, which limits the largest detectable radio features on the sky to approximately $17$ arcmin in size. The data at each frequency was then imaged using multi-frequency-synthesis and uniform weighting at a common pixel scale of $1$ arcsec pixel$^{-1}$ and resolution of $18.0$ arcsec$\times14.8$ arcsec. 

The central frequencies and root-mean-square (RMS) noise values of the images used in further analyses are given in Table~\ref{tab:images}.

\begin{figure}
\begin{center}
\includegraphics[trim=0cm 0cm 0cm 0cm, width=0.495\textwidth]{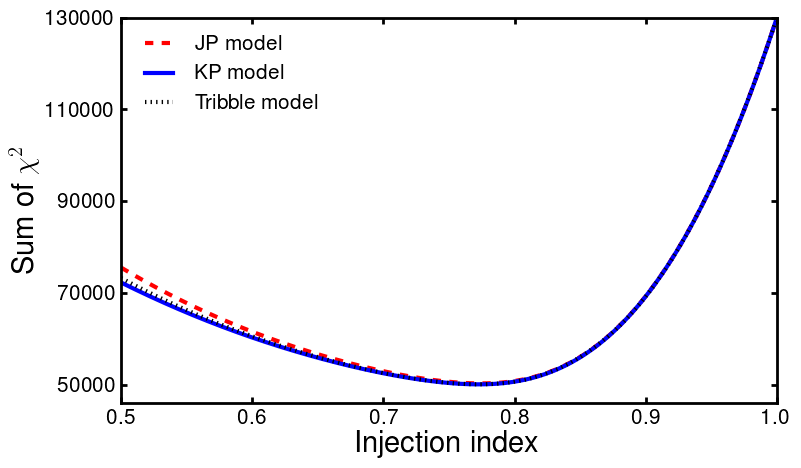}
\end{center}
\caption{$\chi^2$ value for varying injection index for the three injection models fitted with cubic splines. The JP is shown in the red, dashed line, the KP in a solid, blue line, while the Tribble is traced by a dotted, black line.}
\label{fig:injindex}
\end{figure}

\section{Spectral age fitting with {\sc brats}}\label{sec:spec}
As mentioned in Section~\ref{sec:intro}, we have used the {\sc brats} package to fit the JP, KP and Tribble models pixel-by-pixel in six frequency maps. We refer the interested reader to the work by \citet{2013MNRAS.435.3353H} where a description of the {\sc brats} software and its capabilities can be found.

The radio images of the `Sausage' relic were loaded into \textsc{brats}, imposing a $5\sigma$ cutoff level of the radio emission based on the off-source RMS noise (see Table~\ref{tab:images}). The on-source S/N was high enough such that spectra could be fitted on a pixel-by-pixel basis, over $13999$ pixels or an equivalent of almost $50$ independent beams. As shown by \citet{2013MNRAS.435.3353H}, it is of utmost importance to use small regions for fitting synchrotron radio spectra, in order to minimise effects of averaging across electron populations with different ages. 

Spectral ageing models have four free parameters (flux normalisation, magnetic field $B$, injection index $\alpha_\mathrm{inj}$, time since last acceleration $t_\mathrm{age}$). We set the magnetic field input value to $0.6$ nT, as derived by \citet{2010Sci...330..347V}. 

Note that inverse Compton losses because of interactions with the cosmic microwave background (CMB) are important and the code takes these effects into account. This interaction acts as an additional magnetic field $B_\mathrm{CMB}$, added in quadrature to the magnetic field $B$ that permeates the source. 
The CMB energy density in $B$ field terms is \citep{2010hea..book.....L}
\begin{equation}\label{bcmb}
B_\mathrm{CMB}=0.318(1 + z)^2 nT
\end{equation}
where the redshift is $z=0.19$. The equivalent magnetic field for the inverse Compton interactions at $z=0.19$ is $B_\mathrm{CMB}=0.45$ nT.

We fit the JP, KP and Tribble model for a range of injection indices, keeping the flux normalisation and spectral age (time since last acceleration) as free parameters. We perform a broad index injection search between $0.5$ and $1.0$ in steps of $0.1$, followed by a fine search between the initial minima of $0.7$ and $0.8$ in steps of $0.01$. We note that the assumed magnetic field strength of $0.6$ nT \citep[as derived by][]{2010Sci...330..347V} does not affect the measurement of the injection index. 

The goodness of fit (sum of $\chi^2$) dependence on the assumed injection index contains a single minimum (see Fig.~\ref{fig:injindex}). We fix the injection index to the value where the fit reaches the minimum $\chi^2$, i.e. where the model best describes the source brightness distribution. 

An injection index of $\alpha_\mathrm{inj}=0.77$ minimises the $\chi^2$ summed over all the regions for all three models. As shown in Fig. \ref{fig:injindex} the $\chi^{2}$ curve is well known, hence the uncertainty in the injection index can be calculated using the method of \citet{avni76} who shows that the $1\sigma$ error is given by a change in the $\chi^{2}$ value from its minimum of $\Delta \chi^{2} = 1$. However, as we are considering the source on a pixel-by-pixel basis, the sum of $\chi^{2}$ is over weighted by a factor of the beam area $A_\mathrm{beam}$. Assuming the injection index is approximately constant across the source, the errors on the injection index are therefore found by determining a $\Delta \chi^{2}$ of $1$ for the corrected value given by $\chi^{2}_{cor} = \chi^{2}_{min}/A_\mathrm{beam}$. This gives an injection index value for the `Sausage' relic of $0.77^{+0.03}_{-0.02}$. 

\begin{figure*}
\begin{center}
\includegraphics[trim=0cm 0cm 0cm 0cm, width=0.495\textwidth]{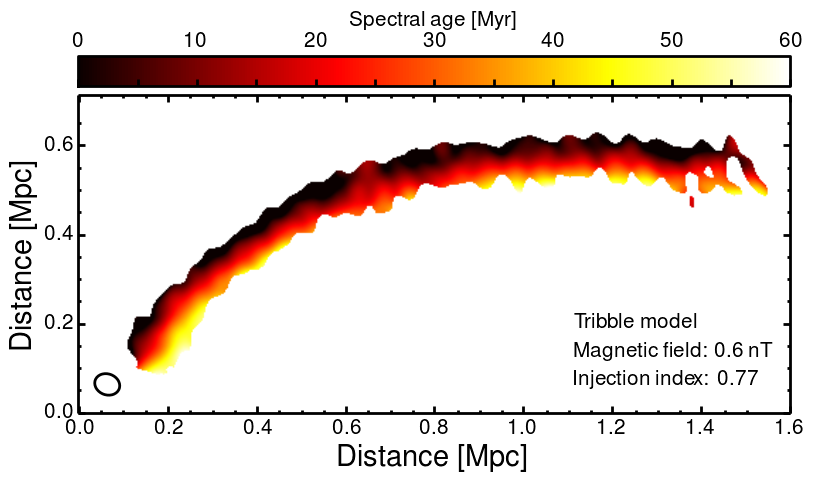}
\includegraphics[trim=0cm 0cm 0cm 0cm, width=0.495\textwidth]{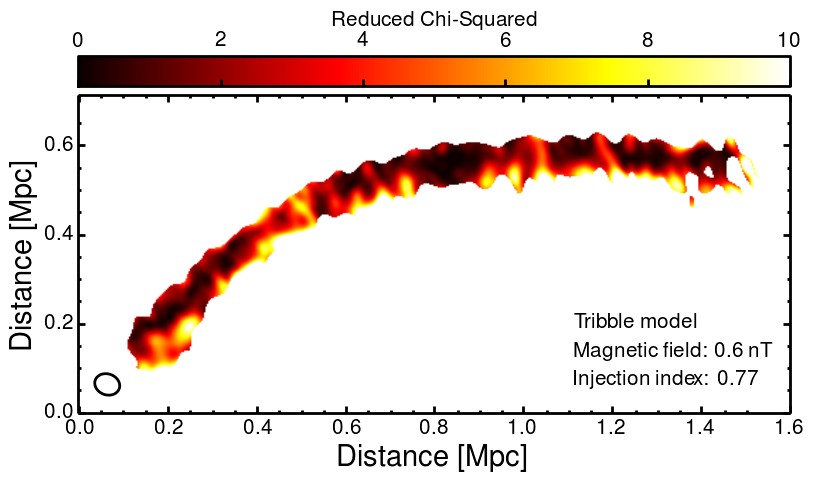}
\end{center}
\caption{Tribble spectral ageing maps to the left and associated $\chi^2$ maps on the right, using a magnetic field value of $0.6$ nT, derived by \citet{2010Sci...330..347V}. The $18.0$ arcsec $\times14.8$ arcsec beam is shown in the bottom-left corner of each image. Note that three models give similar results with a scaling of the ages: the Tribble model predicts the oldest ages, while the KP the youngest.}
\label{fig:ages}
\end{figure*}

We fix the injection index value to $0.77$ and run a final JP, KP and Tribble model fitting to obtain the best-fitting pixel-by-pixel spectral age $t_\mathrm{age}$ and its corresponding $\chi^2$ map. The models have the same goodness-of-fit.  The results for the three models are similar, with variations in ages of the order of $10\%$ (see Fig.~\ref{fig:ages} for the Tribble model results).

\begin{table}
\caption{`Sausage' relic ageing model fitting results. The $0.6$ nT magnetic field has been derived by \citet{2010Sci...330..347V}. The maximum lifetime of electrons is obtained for a $0.26$ nT magnetic field (see Section~\ref{sec:Bfield} for details). `Model' lists the model fitted for the corresponding row. `Mean $\chi^{2}$' lists the average $\chi^{2}$ over the entire source with an equivalent reduced value shown in the `Mean $\chi^{2}_{red}$' column. `Max age' shows the maximum age (time since last acceleration) we could map for each model. Given the S/N cut, we could map emission down to $\sim10\%$ of the peak flux level in the downstream area. The last column represents the error on the maximum age.}
\label{restab}
\begin{tabular}{lcccccc}
\hline
\hline
Model & $B$ field & Mean & Mean & Max Age ($t_\mathrm{age}$) & $\pm$\\
& (nT) & $\chi^{2}$ & $\chi^{2}_{red}$ & (Myrs) & (Myrs)\\
\hline
JP & $0.6$ & 3.60 & 0.90 & 53.3 & 3.7 \\
KP & $0.6$ & 3.58 & 0.90 & 54.2 & 3.7 \\
Tribble & $0.6$ & 3.58 & 0.90 & 59.4 & 3.7 \\
Tribble & $0.26$ & 3.58 & 0.90 & 83.1 & 5.8 \\
\hline
\end{tabular}
\end{table}

\section{Impact of finite resolution}\label{simulations}
We performed simulations to study how the intrinsic spectral ages and the injection index we derive for our `Sausage' data are affected by the convolution of the `true', intrinsic image with the beam of the telescope. 

In order to test this effect, we modelled the `Sausage' relic as a rectangular slab with a thin injection area with a corresponding $0.77$ injection index, followed by JP spectral ageing, according to a shock advance speed of $\sim1500$ km s$^{-1}$ (see Section \ref{sec:discussion} for the derivation of these parameters from the real data). The JP, rather than the KP or Tribble models was used for processing power considerations. The KP model involves the calculation of an extra $\sin^2(\alpha)$ term, compared to the JP. The JP model entails two numerical integrations, and is therefore much faster than the Tribble, which requires an extra numerical integration over the Maxwell-Boltzmann distribution \citep{hardcastle2013}. Maps at the six frequencies ($153$, $323$, $608$, $1221$, $1382$ and $1714$ MHz) were simulated, adding noise as measured in the real GMRT and WSRT maps. The model maps were convolved to $1^2$, $2^2$, $4^2$, $8^2$ and $18.0\times14.8$ arcsec beams. The five sets of maps were run through {\sc brats}, in a similar fashion to the real data. We expect the parameters that we used in the simulations to be representative of the ones recovered from our data, but in any case, no matter which shock speed, magnetic field or model is assumed, the conclusions we draw from our simulations should be similar, apart from a multiplicative factor in the ages determined.

\begin{figure}
\begin{center}
\includegraphics[trim=0cm 0cm 0cm 0cm, width=0.495\textwidth]{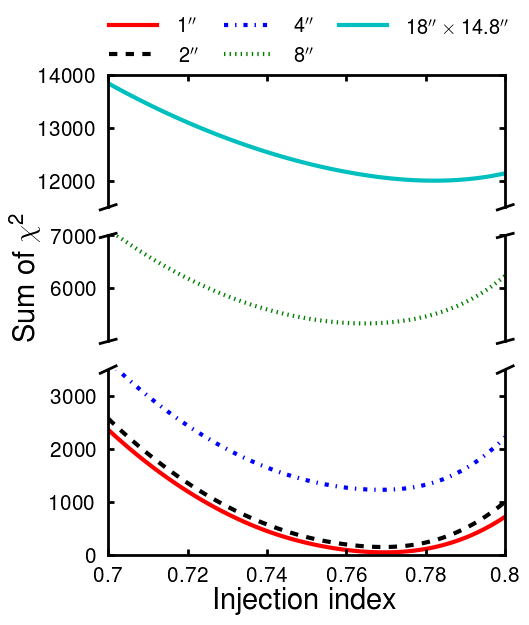}
\end{center}
\caption{$\chi^2$ value for varying injection index calculated for the simulated data, at five different resolutions. All models minimise at a $0.77-0.78$ injection index. The higher resolution maps have lower noise, hence lower $\chi^2$ values.}
\label{fig:chisqsim}
\end{figure}

We searched for the best injection index via $\chi^2$ minimisation and found that the $0.77$ injection index is correctly recovered irrespective of resolution (see Fig. \ref{fig:chisqsim}). The determination of the injection index is dominated by the spectrum at the lowest frequencies, where not much spectral curvature is expected. Moreover, all of the pixels in our radio maps are independently best-fitted by a model traced back to the injection spectrum of $0.77$. We conclude that the injection index estimation in our real GMRT and WSRT maps is robust and consistent with the intrinsic injection index that would be recovered with an infinite-resolution telescope. The injection index can be recovered as long as the ageing, downstream area is still well resolved over at least $2-3$ beams across.

\begin{figure}
\begin{center}
\includegraphics[trim=0cm 0cm 0cm 0cm, width=0.495\textwidth]{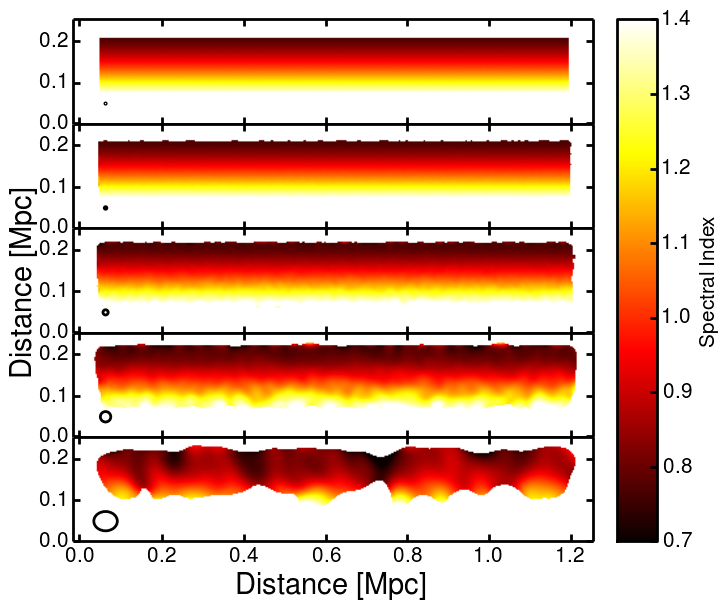}
\end{center}
\caption{Spectral index maps calculated for the simulated data by fitting power laws through the six frequencies. The data are in ascending beam size from top to bottom. The beam is shown in the bottom-left corner of each image.}
\label{fig:spix}
\end{figure}

An important point to make is the noticeable effect of the beam convolution on the spectral index maps (see Fig.~\ref{fig:spix}). Close to the injection area, the correct index is recovered in the higher resolution maps (up to $2$ arcsec). As the beam size increases, the injection area is not resolved and the mixing of noise with fresh plasma leads to a flattening of the measured spectral index, up to $\sim0.6$, in parts of the injection area. A noise maximum or minimum on top of radio emission in one of the radio maps can lead to an over-estimation or under-estimation of the spectral index within that region. For example, if our lower frequency map suffers from a negative noise peak this artificially flattens the radio spectrum. The mixing of real emission with noise right in front of the shock, depending on the noise properties in each map, can also lead to a flattening of the measured injection index. This then naturally explains what is seen in the real `Sausage' data, where the spectral modelling finds a best injection index of $0.77$, while in the spectral index maps we measure values of about $0.60-0.65$ \citep{2013A&A...555A.110S}. All previous studies of radio relics \cite[e.g. MACSJ1149.5+2223, MACSJ1752.0+4440, Abell 521, Abell 2744, Abell 754, CIZA J2242.8+5301, 1RXS J0603.3+4214;][]{2007A&A...467..943O, 2008A&A...486..347G, 2010Sci...330..347V, 2011ApJ...728...82M, 2011MNRAS.418..230V, 2012MNRAS.426...40B, 2013A&A...555A.110S} use the flattest index measured in the spectral index map as the injection index or calculate the injection index from the integrated spectral index. The integrated spectral index of the `Sausage' relic has been shown to vary significantly depending of the observed frequency window \citep{AMI}. We therefore caution against using the spectral index map or integrated index as a proxy for injection index determination.

\begin{figure*}
\begin{center}
\includegraphics[trim=0cm 0cm 0cm 0cm, width=0.495\textwidth]{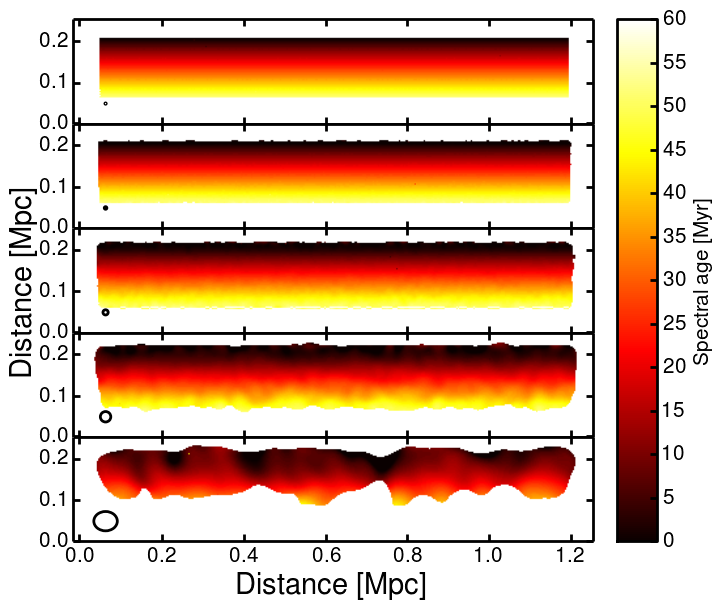}
\includegraphics[trim=0cm 0cm 0cm 0cm, width=0.495\textwidth]{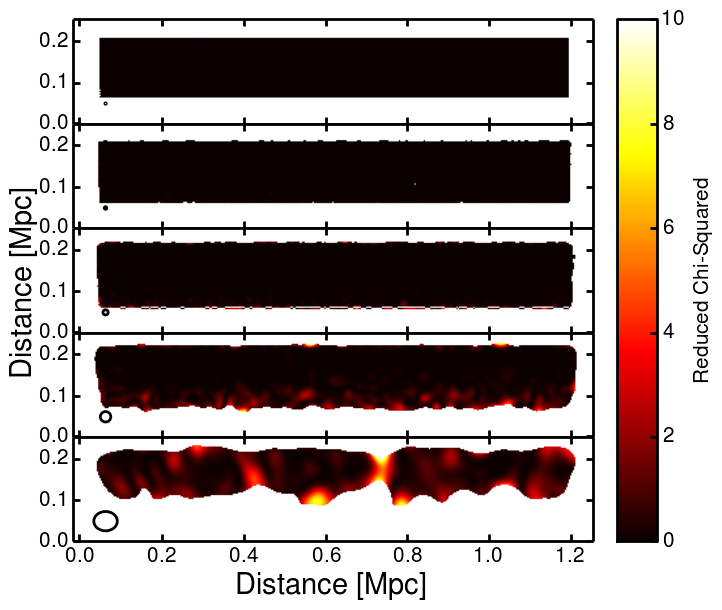}
\end{center}
\caption{JP spectral ageing maps to the left and associated $\chi^2$ maps on the right for the simulated data. The data are in order of ascending beam size from top to bottom: $1^2$, $2^2$, $4^2$, $8^2$, $18.0\times14.8$ arcsec. The beam is shown in the bottom-left corner of each image.}
\label{fig:agessim}
\end{figure*}

We also tested the effect of resolution on the age determination (see Fig. \ref{fig:agessim}).
Decreasing the resolution of the radio data leads to mixing of different electron populations within an observed beam. With a decrease of resolution, the low-frequency spectrum index of each given region within the relic gets steeper, while its high-frequency index is measured as flatter, pushing the radio spectrum to a power-law. Therefore, this is equivalent to underestimating the curvature of the spectrum at high frequencies. While for lower-resolution images, the injection index derived is the same as in the higher-resolution maps, there is an effect of overestimating the spectral age, hence underestimating the spectral speed. For example, if we decrease the resolution from $1^2$ arcsec to $18.0\times14.8$ arcsec, the derived age increases by $\sim10$ per cent. These simulations at the lowest resolution, $18.0\times14.8$ arcsec, are comparable to our real GMRT and WSRT maps. $10$ per cent is well within our measurement error in the real maps. Therefore resolution effects will not play a major role in our age measurements.

\section{Discussion}
\label{sec:discussion}

\subsection{Best-fitting model?}\label{sec:bestmodel}
We fit the JP and KP models of spectral ageing, along with the more recent Tribble model directly to six radio-frequency data of the well-studied `Sausage' relic, to derive a consistent set of shock parameters.

The three models have similar goodness of fit, with a mean $\chi^2=3.6$. All three models fit the data equally well. Nevertheless, we would expect, given the well-ordered magnetic fields measured in proximity to the `Sausage' shock front, that the JP or KP would provide the better fit. Since the constant pitch angle of the KP is a less physically plausible scenario, the JP would be the preferred model. \citet{2013A&A...555A.110S}, by using a colour-colour analysis, concluded that the best description of the data was given by a superposition of JP models with different age. However, the comparison to the spectral ageing models was done on large areas, summing up many pixels with similar spectral index properties to increase S/N. This techniques does not allow the probing of the extremely steep spectrum plasma in the downstream area which would have allowed conclusive discrimination between the JP and KP model \citep[the Tribble model was not tested by ][]{2013A&A...555A.110S}.

The source is highly-polarised ($60$ per cent) towards the location of the shock \citep{2010Sci...330..347V}, but this is not necessarily true for the much fainter downstream area. The measurement of the polarisation is dominated by the high surface brightness plasma close to the shock front. The amplification and alignment of the magnetic field vectors is likely a result of shock compression \citep[e.g][]{2012MNRAS.423.2781I}. While this is a reasonable assumption close to the shock front, in the downstream area there is the possibility of highly-turbulent magnetic fields, which motivates the use of a Tribble-like model. Steepening of the integrated relic spectrum towards $16$ GHz also supports a scenario which includes either turbulent magnetic fields or ICM inhomogeneities \citep{AMI}. 

In further analyses, parameters will be derived according to the best-fitting Tribble model.

\subsection{Injection index}\label{sec:injindex}
Assuming a Tribble model, we will derive a number of physical parameters for the `Sausage' shock. The  injection index $0.77^{+0.03}_{-0.02}$ we calculate is significantly steeper ($6\sigma$ away) than what was found before from spectral index mapping and colour-colour plots \citep[$\alpha_\mathrm{inj}=0.65\pm0.05$;][]{2010Sci...330..347V,2013A&A...555A.110S}. The discrepancy between the steep injection index derived from spectral modelling and the old measurement can be explained by convolution effects, as shown in Section \ref{simulations}.

\subsection{Mach number}\label{sec:mach}
The Mach number $M$ of a shock moving through a medium at speed $v_\mathrm{shock}$ is defined as
\begin{equation}\label{eq:mach}
M=\frac{v_\mathrm{shock}}{c_s}
\end{equation}
where $c_\mathrm{s}$ is the sound speed in the upstream medium. 

For simple shocks, the Mach number can be related to the radio injection index by \citep{1987PhR...154....1B}
\begin{equation}\label{machnumber}
M=\sqrt{\frac{2\alpha_\mathrm{inj}+3}{2\alpha_\mathrm{inj}-1}}
\end{equation}
Hence, for $\alpha_\mathrm{inj} = 0.77^{+0.03}_{-0.02}$ (see Sect \ref{sec:injindex}), we derive a Mach number of $2.90^{+0.10}_{-0.13}$ for the relic.

A number of papers based on \textit{XMM-Newton} and \textit{Suzaku} data \citep{2013PASJ...65...16A, 2013MNRAS.429.2617O, 2013MNRAS.433..812O, 2014MNRAS.440.3416O} reported a possible discrepancy between the X-ray and the radio shock Mach numbers \citep[e.g.][]{2010Sci...330..347V,2012A&A...546A.124V,2013A&A...555A.110S}. In the case of the `Sausage' relic, the X-ray Mach number based of Suzaku data is $2.54^{+0.64}_{-0.43}$ \citep{2014MNRAS.440.3416O}. Hence, the Mach number derived from the X-ray data also favours a steep injection index, close to $0.8$ (via equation \ref{machnumber}).

The weaker shock strength found here using spectral fitting would reconcile the differences found between radio and X-ray derived shock parameters for this radio relic. DSA happens to all charged, high energy particles when they encounter a collisionless shock too thin to capture them and depends only on the particle rigidity (momentum per charge). Simple DSA assumes that thermal particles are injected. Because of the much higher rigidity of thermal protons compared to thermal electrons, the injection efficiency of thermal protons is much higher than that of thermal electrons. For example, the electron efficiency for strong shocks ($M\gg10$) is much lower than that of protons, at a ratio of $\sim0.001$ \citep{2014IJMPD..2330007B}. If such a low DSA electron acceleration efficiency is assumed, the weak  Mach number derived in this paper challenges even more the simple DSA theory of thermal particle acceleration proposed for radio relics. In the case of weak shocks ($M\ll10$), the electron efficiency is not well constrained \citep{2014IJMPD..2330007B}. Simple DSA could in principle be reconciled with a low Mach number, given a very high electron efficiency, close to that of protons. Alternative theories including an injection of pre-accelerated rather than thermal particles \citep[e.g.][]{2012ApJ...756...97K}, turbulent re-acceleration \citep[pre-accelerated relativistic particles upstream can be maintained by several mechanisms including turbulent re-acceleration, e.g.][]{2001MNRAS.320..365B} or intermittent/highly turbulent magnetic fields could also explain the high level of radio emission, even with a low Mach number shock \citep[e.g.][]{2012MNRAS.423.2781I}. Our results also seem to support the simulations by \citet{2008ApJ...689L.133B} and \citet{2012ApJ...756...97K}, which predict that a $M\approx2$ shock could inject enough particles to reproduce the level of radio emission observed in the Sausage cluster, given a pre-existing low-energy, cosmic ray electron population.

\subsection{Shock and ageing speed from spectral modelling}\label{sec:v}
The shock advance speed can be derived from the Mach number, in two independent ways, by using radio and X-ray data.

The X-ray ICM gas temperature in the upstream of the `Sausage' shock front has been measured by \citet{2014MNRAS.440.3416O} using \textit{Suzaku} to be $3.35^{+1.13}_{-0.72}$ keV. Using the conversion $c_\mathrm{s}=1480 \sqrt{T/(10^8K)}$ km s$^{-1}$ from \citet{1986RvMP...58....1S}, this results in a sound speed $c_\mathrm{s} = 920^{+150}_{-110}$ km s$^{-1}$ . The Mach number derived from the X-ray temperature data is: $M=2.54^{+0.64}_{-0.43}$ \citep{2014MNRAS.440.3416O}, thus a shock advance speed of $v_\mathrm{shock}=2340^{+700}_{-490}$ km s$^{-1}$.

A second measure of the shock speed can be obtained from the particle acceleration analysis. The new radio Mach number of $M=2.90^{+0.10}_{-0.13}$ constrains the shock speed to $v_\mathrm{shock}=2670^{+450}_{-340}$ km s$^{-1}$.

With the injection index and X-ray Mach numbers being in agreement, we have two independent measurements of shock advance speeds in the context of cluster shock waves. We place the preferred shock speed $v_\mathrm{shock}$ at the average value between these two methods at $\approx2500$ km s$^{-1}$.

The spectral ageing data can also be used to derive a measurement of the ageing speed $v_\mathrm{age}$, the speed at which the particles in the downstream area of the shock flow away from the shock, in the shock reference frame.

Standard shock physics says \citep{1987PhR...154....1B}
\begin{equation}
v_\mathrm{age}/v_\mathrm{shock}=\frac{\gamma-1+2/M^2}{\gamma+1}
\end{equation}
where $v_\mathrm{age}$ is the ageing speed (the speed at which the plasma flows away from the shock), $v_\mathrm{shock}$ is the shock upstream of the shock, i.e. the shock advance speed itself. $\gamma=5/3$ is the ratio of specific heats and $M$ is the Mach number. Therefore, the ageing speed is
\begin{equation}\label{eq:vage}
v_\mathrm{age}=c_s \frac{\gamma M -M +2/M}{\gamma+1}.
\end{equation}
Given the Mach number derived above ($M=2.90$) and $c_s=920$ km s$^{-1}$, this leads to an expected ageing speed $v_\mathrm{age}=905^{+165}_{-125}$ km s$^{-1}$.

\subsection{Spectral age}\label{ages}
In this section we consider the constraints on spectral age and magnetic field strength imposed by the observations. 

\subsubsection{Spectral age}
Given the newly determined injection index ($0.77$) and assuming a magnetic field strength of $0.6$ nT, we can derive spectral ages using the Tribble model (as explained in Sec \ref{sec:bestmodel}).

We see clear ageing from north to south across the source, i.e. from the front of the shock front into the downstream area (left column of Fig.~\ref{fig:ages}). The age $t_\mathrm{age}$ (time since initial acceleration) increases systematically across the source, with pixels located at the same projected distance from the expected location of the shock front having the same spectral age, meaning they were last accelerated at the same time. The ages along the length of the source are very similar at the same distances from the location of the shock (variations given by the noise are below $9$ per cent). The age depends linearly on the distance from the shock front, indicating that the shock speed has stayed constant for the past tens of Myr. 

With a magnetic field of $0.6$ nT, the maximum age is $\sim60$ Myr (see Fig \ref{fig:ages}). There are some age variations that come from small-scale, beam-size age variations across the length of the source (maximum of $5$ Myr), which coincide with noise peaks in the $153$ MHz map. This was also noted in the context of spectral index maps by \citet{2013A&A...555A.110S}. By excluding these peaks, we can see that the spectral speed varies little along the length of the relic, with a standard deviation of $300$ km s$^{-1}$. To increase S/N, we average pixels along concentric circles fitted to describe the shape of the outer edge of the relic. The dependence of the average age with distance from the location of the shock front can be visualised in Figure~\ref{fig:age_profile}. The typical standard deviation of pixels along these concentric circles is $\sim3$ Myr (per pixel). Assuming constant ICM temperature along the relic length, we can expect variations of at most $10$ per cent in Mach number along the source.

While spectral index maps provide a proxy for spectral ageing, spectral steepening could be seen in some radio relics as a result of fortuitous superposition of filamentary radio structure. By using hydrodynamical simulations of cluster mergers, \citet{2013ApJ...765...21S} argue that, even without prescribing spectral ageing, it is possible to observe spectral index trends caused by unrelated shock structures with different Mach numbers, seen in fortuitous projection at the same location. The authors therefore suggest that even the most basic assumptions of the formation mechanism of radio relics (shock-accelerated particles, with subsequent ageing) do not hold. In our analysis, we present the first direct measurement of aged plasma in the context of radio relics. We would like to stress that the age measurements are independently fitted on a pixel-by-pixel basis, therefore the systematic increase which is consistent across the entire $1.4$ Mpc length of the source is undoubtedly a real effect, which cannot come from superposition of unrelated shock structures.

\subsubsection{Magnetic field -- constraints on spectral age}\label{sec:Bfield}
Spectral ageing models have four free parameters: flux normalisation, injection index $\alpha_\mathrm{inj}$, spectral age $t_\mathrm{age}$ and magnetic field strength $B$. Given a fixed flux normalisation $F$ and injection index $\alpha_\mathrm{inj}$, to reproduce a given spectral shape at the high-frequency end, where losses dominate, the age of the electrons relates to the magnetic field in the following way \citep{1993MNRAS.261...57T, hardcastle2013}:
\begin{equation}
\label{eq:t_B}
t_\mathrm{age} = C(\alpha_\mathrm{inj}, F) \frac{\sqrt{B}}{B^2+B^2_\mathrm{CMB}}
\end{equation}
where $C(\alpha_\mathrm{inj}, F)$ is a constant that depends on the fixed flux normalisation and injection index, $B$ is the source magnetic field and $B_\mathrm{CMB}=0.45$ nT is the equivalent magnetic field exerted by the inverse Compton interactions at $z=0.19$. By setting the derivative of equation \ref{eq:t_B} to 0, we can calculate the maximum lifetime of electrons $t_\mathrm{max}$ is obtained for a magnetic field
\begin{equation}
B = B_\mathrm{CMB}/\sqrt{3}
\end{equation}
Equation \ref{eq:t_B} becomes:
\begin{equation}
t_\mathrm{max} \propto \frac{\sqrt{B_\mathrm{CMB}}}{\frac{4}{3}\sqrt[4]{3}B^2_\mathrm{CMB}}
\end{equation}
The age obtained from the Tribble model fit with a $0.6$ nT magnetic field at $10\%$ flux level in the downstream area, or $\sim150$ kpc away from the shock location, is $59.4$ Myr. $t_\mathrm{age}$ within this equation has a clear maximum, attained for a magnetic field of $0.26$ nT. By scaling the age using this magnetic field value, the maximum lifetime of electrons in the Tribble model is $\sim83.1$ Myr, at a distance of $\sim150$ kpc from the shock front. 

\subsubsection{Spectral age discrepancy}
\begin{figure}
\begin{center}
\includegraphics[trim=0cm 0cm 0cm 0cm, width=0.495\textwidth]{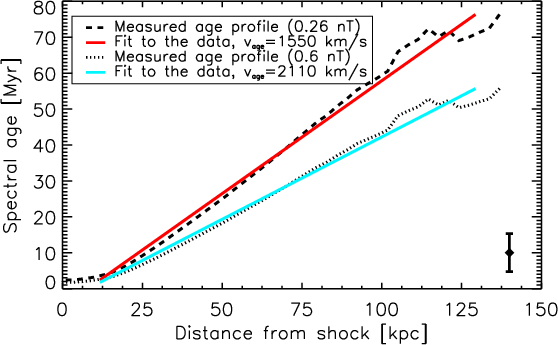}
\end{center}
\caption{Spectral age profile as function of distance from the shock front. The age increases systematically with distance from the shock. In the scenario where the electrons are accelerated by the shock front located at the north of the relic, for a minimum $B$ field value of $0.26$ nT, yielding a maximum electron lifetime, a linear model yields a speed $v_\mathrm{age}$ of $1550$ km s$^{-1}$. For a typical magnetic field measured in radio relics ($\sim0.6$ nT), the measured speed is higher ($2210$ km s$^{-1}$). We obtain the profile by averaging ages along the length of the relic. The typical standard deviation of each pixel-based average age value (for $0.26$ nT) is shown in the bottom-right corner. The allowed spectral ages are too young to explain the very slow ageing speed imposed by the shock Mach number and upstream sound speed (see Section~\ref{sec:Bfield}).}
\label{fig:age_profile}
\end{figure}

Our model fitting is sensitive to the time since the electrons population last underwent acceleration. Assuming the last agent of acceleration was the Sausage shock now located at the northern edge of the relic, the electrons should age freely in the magnetic field after the shock passage. The ageing speed $v_\mathrm{age}$, or speed at which the electrons move away from the `Sausage' shock in this scenario, is directly determined by $t_\mathrm{age}$ and the distance from the shock front. By taking into account the age dependence $t_\mathrm{age}$ on the distance $d$ from the shock front, we can define an ageing speed $v_\mathrm{age}=d/t_\mathrm{age}$ (see Figure~\ref{fig:age_profile}). At the distance of $150$ kpc, the maximum lifetime of electrons is $83.1$ Myr, hence the minimum ageing speed is $\sim1550$ km s$^{-1}$. For any B field weaker or stronger than $B_\mathrm{CMB}/\sqrt{3}$, the ageing speed must be higher (e.g. for $0.6$ nT the speed rises to $2110$ km s$^{-1}$).

Therefore, from our data, we can find a minimum on the speed at which the electrons can flow away from the shock front. The minimum ageing speed constrained here from the B field is $v_\mathrm{age}>1550$ km s$^{-1}$. Hence, the relationship between the shock speed and the ageing speed imposes $v_\mathrm{age}=905^{+165}_{-125}$ (derived in Section~\ref{sec:v}). These two derivations of the ageing speed are incompatible at $>3\sigma$ level. This discrepancy suggests that the electrons are ageing too slowly given their distance from the northern edge of the `Sausage' relic, in a scenario where the particles are last accelerated by the `Sausage' shock front (see also Fig.~\ref{fig:age_profile}).

\subsubsection{Magnetic fields or turbulence?}
In the spectral ageing models, magnetic field and ageing lifetime are interchangeable, as shown in Section \ref{sec:Bfield}. Therefore, we could bring the two ageing speeds into agreement by more closely inspecting these two parameters.

The magnetic field imposes a maximum electron lifetime, but from other constrains we expect the magnetic field to be significantly higher than the maximum electron lifetime B field of $0.26$ nT. Estimates from simulated radio relic profiles from \citet{2010Sci...330..347V} indicate a B field of $0.5-0.7$ nT. Rotation measure synthesis at similar cluster-centric distances for other clusters indicates values of $0.01-0.5$ nT \citep[e.g.][]{2001ApJ...547L.111C,2006A&A...460..425G,2011A&A...530A..24B}. $B$ can be increased by simple compression at the shock by up to the density compression factor (in the limit of a perpendicular shock), which is $\sim 2.4$ for the Mach number derived above; the fact that the Faraday-rotation-corrected $B$-field vectors are perpendicular to the shock \citep{2010Sci...330..347V} is consistent with a compression model, where the shock front magnifies and aligns the magnetic field vectors in its vicinity. Numerical modelling \citep[e.g.][]{2012MNRAS.423.2781I} shows magnetic field amplification factors consistent with this in more realistic field configurations; it is possible that cosmic-ray-mediated magnetic field amplification, which would give a field increase over and above simple compression, occurs in relics in the same way as in supernova remnants \citep[e.g.][]{2005A&A...433..229V}. We have few other direct constraints on the field strength in the relic region. We have estimated from the {\it Chandra} data described by \citet{2014MNRAS.440.3416O} an upper limit on the inverse-Compton emission which corresponds to a lower limit on the field strength of around $0.1$ nT: this is extremely conservative because it attributes all the net X-ray emission from the relic region to the inverse-Compton process, whereas we would expect some of it to be thermal. The equipartition field strength in the relic region (assuming that the injection index that we measure extends to low energies) would be around $0.3$ nT. Equipartition assumes that the energy is split between magnetic field energy and energy of the synchrotron electrons. This neglects the contribution to the energy density of the thermal protons and electrons, which is almost certainly dominant, so we can expect the real field strength to be higher than the equipartition value. Thus we conclude that the discrepancy between the two ageing speed determinations is likely even more substantial than what the limit on speed would indicate.

While testing the established JP, KP and Tribble models is a natural step forward from traditional techniques of spectral index maps and colour-colour plots, the ageing results are internally inconsistent, as shown above. There are a few effects that could ameliorate the inconsistency between the ageing solution and the shock speed. The following processes would affect the values of the ages derived in the downstream area of the shock.

Part of the discrepancy could be removed if, after the passage of the `Sausage' shock front, there is another source of re-acceleration of the particles, which would bring the radio spectrum closer to an injection spectrum. However, re-acceleration would not necessarily allow the low-frequency, injection spectrum to remain unchanged (thus not preserving the measured injection index). Possible sources of re-acceleration in the case of radio relics could be through turbulence in the downstream area of the `Sausage' shock or minor shock/instabilities created in the downstream area after the passage of the main shock front. Turbulence is also thought to be the main driver of re-acceleration within the ICM of merging clusters, leading to the formation of radio haloes, diffuse radio emission co-spatial with the X-ray emitting gas of some merging, luminous clusters \citep[e.g.][]{2001MNRAS.320..365B}. Small-scale turbulence would allow re-acceleration of particles, mimicking a spectral age lower than what would be expected from the distance to the `Sausage' shock front, without leading to too much mixing of electron populations across the width and length of the relic. This way, the age gradient we observe across the radio relic could be preserved. Significant turbulence is expected in the downstream of merger shocks from simulations \citep[e.g.][]{2012MNRAS.423.2781I}. Higher-frequency, higher-resolution observations ($2-5$ GHz Very Large Array observations, van Weeren et al. in prep) would provide a good test for this theory, where spectral index fine structure across the relic is expected. Higher spatial resolution maps would also show saturating spectral index values with distance from the shock, as turbulence would overcome the ageing as soon as the ageing time approaches the turbulence acceleration time. However, there are challenges to this scenario. Firstly, the spectral evolution of a population of electrons undergoing re-acceleration via turbulence would most likely be different than what is expected from the JP, KP or Tribble models, which makes the good fits that we obtain using these models puzzling. Secondly, in the standard scenario only a relatively small fraction of the turbulent energy that is transported from large to small scales is channelled into electron re-acceleration \citep{2007MNRAS.378..245B}, with the result that the acceleration time-scales would be of the order of $100$ Myrs, significantly larger than those needed in the `Sausage' relic.  Given its low Mach number, the `Sausage' shock would not generate enough turbulent energy to re-accelerate particles to speeds beyond a few hundreds of km s$^{-1}$ on small scales ($\ll100$ Myr). Boosting the acceleration efficiency in turbulent models requires special circumstances, including a reduced effective mean-free-path in the ICM \citep{2011MNRAS.410..127B} or an efficient generation of Alfv{\'e}n waves at quasi-resonant scales \citep{2002ApJ...577..658O,2004MNRAS.350.1174B}.

A second possibility is fast (super-Alfv{\'e}nic) particle diffusion in the downstream area. However, this scenario requires very special  conditions, such as highly intermittent magnetic fields (as in magnetic reconnecting regions) and/or an efficient damping of plasma instabilities \citep{2013MNRAS.434.2209W,2014IJMPD..2330007B}.

Another point to note is that mixing along the line of sight will lead to the measurement of younger ages than expected given the distance from the shock front. In a simplified scenario, the `Sausage' shock is a spherical cap front propagating northwards, seen in projection on the sky. Along any given line of sight, emission with a range of spectral ages actually located at different distances from the shock front is projected at the same apparent distance from the shock front. Mixing is expected to have the strongest effect furthest away from the shock, in the downstream area, along the merger axis of the cluster. The detailed arc-length and further 3D structure of the shock has been shown to have a significant effect on the surface brightness and spectral properties \citep{2012ApJ...756...97K}. We will explore the effects of this mixing on radio spectra in a later paper.

To summarize, all the effects we have considered above have difficulties in explaining the observations and none is expected to solely bridge the discrepancy between the ageing speed as derived from the spectral age and shock speed data. Further modelling is necessary to fully investigate the nature of the ageing behind the shock.

\subsection{Merger history}
Assuming a constant shock speed throughout the travel towards the outskirts of the cluster, we can trace back the formation time of the shock. We expect the shock speed to have been slightly different in the past given its dependence on the ICM temperature, density and pressure, which can vary across the cluster in a complex way, given how disturbed the cluster is \citep[see][for temperature and density maps of the cluster]{2013MNRAS.433..812O}. By fixing the cluster centre, from X-ray \citep[emission peak,][]{2014MNRAS.440.3416O} or weak lensing (mass distribution, Jee et al. in prep) arguments, we can calculate the time required for the gravitationally-decoupled shock front to travel from the core-passage of the two merging sub-clusters. In the case of the `Sausage' cluster, this analysis indicates that the time since core passage is $\sim0.6-0.8$ Gyr. Albeit a very approximate value which should be considered with caution, the time since core passage is consistent with the results from a much more sophisticated Monte-Carlo dynamical analysis which will be presented in Dawson et al. (in prep), based on the method from \citet{2013ApJ...772..131D}. 

\section{Conclusions}
\label{sec:conclusion}
In this paper, we have performed the first spectral age model fitting of a radio relic. Multi-frequency GMRT and WSRT radio data spanning six equally-spaced bands between $153$ and $1714$ MHz enabled us to draw a number of conclusions. We have derived injection indices and spectral ages for the plasma using established electron injection models such as the JP, KP and Tribble models. 
\begin{itemize}
\item Even with the best radio relic data available, we could not distinguish between the three spectral ageing models. Wider-bandwidth, higher-frequency and more reliable flux measurement would help distinguishing between ageing models.
\item An injection spectral index of $0.77^{+0.03}_{-0.02}$ was derived, steeper than previously computed from radio data. The differences can be reconciled when taking beam smearing effects into account, which affect the old spectral index derivation only. We calculate a secure Mach number of $2.90^{+0.10}_{-0.13}$, based on a shock-acceleration model.
\item We find a systematic increase in plasma age within the downstream area at the back of the shock front and uncover old plasma, with ages up to $60$ Myr, assuming a $0.6$ nT magnetic field.
\item The new radio-derived Mach number is consistent with the X-ray results of \citet{2014MNRAS.440.3416O}, which imply a Mach number of $2.54^{+0.64}_{-0.43}$.
\item We derive shock advance speeds of $\sim2500$ km s$^{-1}$ from X-ray data and radio particle acceleration arguments.
\item The spectral ages allowed by a model of freely-ageing electrons behind the `Sausage' shock front and the ages expected given the shock speed are incompatible. Part of the discrepancy between the two measurements can be attributed to line-of-sight mixing. They could be further brought into agreement if subsequent re-acceleration happens after the passage of the main shock front or if there is an usually efficient diffusion of particles in the downstream area.
\end{itemize}
The results presented here act as a proof-of-concept to display the power of multi-frequency data and to strengthen the necessity for broad-band observations of radio relics. Such observations for other clusters hosting radio relics could pose tight constrains on the clumpiness of the ICM. Very low frequency LOFAR measurement are crucial for the accurate determination of the injection index and Mach number. 

\section*{Acknowledgements}
We would like to express our gratitude to our referee, Thomas Jones, who provided us with detailed comments which improved the scientific interpretation of our results. We would like to especially thank Marcus Br\"uggen and Gianfranco Brunetti for their valuable theoretical input. We also thank Walter Jaffe, George Miley, Leah Morabito, Will Dawson, David Sobral, David Wittman, Vera Margoniner and Reinout van Weeren for useful discussions. AS acknowledges financial support from NWO. JJH thanks the STFC for a STEP award and the University of Hertfordshire for generous financial support. We thank the staff of the GMRT who have made these observations possible. The GMRT is run by the National Centre for Radio Astrophysics of the Tata Institute of Fundamental Research. The Westerbork Synthesis Radio Telescope is operated by the ASTRON (Netherlands Institute for Radio Astronomy) with support from the Netherlands Foundation for Scientific Research (NWO). This research has made use of the NASA/IPAC Extragalactic Database (NED) which is operated by the Jet Propulsion Laboratory, California Institute of Technology, under contract with the National Aeronautics and Space Administration. This research has made use of NASA's Astrophysics Data System. 

\bibliographystyle{mn2e.bst}
\bibliography{Sausage_BRATS}

\label{lastpage}
\nocite{*}
\end{document}